\documentclass[letterpaper,twocolumn,american,showpacs,prb,aps,superscriptaddress]{revtex4}
\usepackage[latin1]{inputenc}
\usepackage{bm}
\usepackage{multirow,amssymb,amsbsy,amsmath}
\usepackage{stmaryrd}
\usepackage{graphicx}
\usepackage{color}
\makeatletter
\usepackage{pifont}
\makeatother \DeclareMathOperator{\Tr}{Tr}

\begin{document}

\title{Non-Markovian dynamics of a
       single electron spin coupled to a nuclear spin bath}

\author{E. Ferraro}

\email{ferraro@fisica.unipa.it}

\affiliation{Dipartimento di Scienze Fisiche ed Astronomiche,
Universit\`a di Palermo, via Archirafi 36, 90123 Palermo, Italy}

\author{H.-P. Breuer}

\email{breuer@physik.uni-freiburg.de}

\affiliation{Physikalisches Institut, Universit\"at Freiburg,
             Hermann-Herder-Strasse 3, D-79104 Freiburg, Germany}

\affiliation{Hanse Wissenschaftskolleg, Institute for Advanced
             Study, D-27733 Delmenhorst, Germany}

\author{A. Napoli}

\affiliation{Dipartimento di Scienze Fisiche ed Astronomiche,
Universit\`a di Palermo, via Archirafi 36, 90123 Palermo, Italy}

\author{M. A. Jivulescu}

\affiliation{Dipartimento di Scienze Fisiche ed Astronomiche,
Universit\`a di Palermo, via Archirafi 36, 90123 Palermo, Italy}

\author{A. Messina}

\affiliation{Dipartimento di Scienze Fisiche ed Astronomiche,
Universit\`a di Palermo, via Archirafi 36, 90123 Palermo, Italy}

\date{\today}

\begin{abstract}
We apply the time-convolutionless (TCL) projection operator
technique to the model of a central spin which is coupled to a
spin bath via nonuniform Heisenberg interaction. The second-order
results of the TCL method for the coherences and populations of
the central spin are determined analytically and compared with
numerical simulations of the full von Neumann equation of the
total system. The TCL approach is found to yield an excellent
approximation in the strong field regime for the description of
both the short-time dynamics and the long time behavior.
\end{abstract}


\maketitle

\section{Introduction}

Projection operator techniques \cite{KUBO} are widely used in
studies of the dynamical behavior of complex open quantum systems
featuring non-Markovian relaxation and decoherence phenomena
\cite{Breuer1}. The most prominent variant of these techniques is
the Nakajima-Zwanzig (NZ) projection operator method which leads
to an integrodifferential equation for the reduced density matrix
of the open system containing a certain memory kernel
\cite{NAKAJIMA,ZWANZIG}. An alternative and technically much
simpler scheme is the time-convolutionless (TCL) projection
operator technique in which one obtains a first-order differential
equation for the reduced density matrix \cite{SHIBATA}. The
advantage of the TCL approach consists in the fact that it yields
an equation of motion for the relevant degrees of freedom which is
local in time and which is therefore often much easier to deal
with than the NZ master equation. In fact, this method has been
applied to many physical systems showing strong non-Markovian
effects (see, e.~g., Refs.~\onlinecite{DECOHERENCE,AHN02,BGM}).

In the present paper we apply the TCL projection operator
technique to the model of a central spin interacting with a bath
of $N$ spins defined by the Hamiltonian ($\hbar$=1)
\begin{equation} \label{Hamiltonian}
 H = \frac{\omega_0}{2}\sigma_3 +
 \sum_{k=1}^N \alpha_k \, \bm{\sigma} \cdot \bm{\sigma}^k.
\end{equation}
The Pauli operators $\bm{\sigma}$ and $\bm{\sigma}^k$ act on the
Hilbert spaces of the central spin, which is regarded as the open
quantum system, and of the $k$-th bath spin, respectively. The
strength of the spin-bath coupling is given by the constants
$\alpha_k$. Moreover we have included an external magnetic field
that acts on the central spin and leads to the Zeeman splitting
$\omega_0$.

The model given by the Hamiltonian (\ref{Hamiltonian}) may be used
to describe for example a single, localized electron spin coupled
to a bath of nuclear spins in a quantum dot through contact
hyperfine interaction \cite{Schliemann}. It features many
interesting phenomena such as non-exponential behavior of
correlations and coherences and strong non-Markovian effects. A
detailed treatment of the model within the NZ projection operator
technique has been carried out in Ref.~\onlinecite{Coish}, and
non-perturbative solutions for polarized initial conditions have
been constructed in Ref.~\onlinecite{Khaetskii}. Recently, a
detailed analytical and numerical study of the exact Bethe ansatz
solution \cite{Gaudin} of the model has been carried out
\cite{Bortz}. Moreover, several efficient numerical algorithms
have been proposed that are based, e.~g., on the
spin-coherent-state representation \cite{Harmon} or on the
Chebyshev expansion for the full propagator \cite{DeRaedt}.

The TCL projection operator method has been applied to various
spin bath models for which a compact analytical solution is
available, such as central spin models with Heisenberg XY
\cite{BBP} and with full Heisenberg interaction \cite{Fischer} for
uniform couplings, and central spin models with nonuniform Ising
interaction \cite{LIDAR07}. The purpose of the present paper is a
detailed investigation of the performance of the TCL technique for
the nontrivial model given by Eq.~(\ref{Hamiltonian}) with
nonuniform couplings. To this end, we will compare the results for
the populations and the coherences of the central spin obtained
from the TCL approach with numerical simulations of the full von
Neumann equation of the model. It will be demonstrated that the
method provides an efficient scheme which is applicable in the
perturbative regime of weak couplings, even for long interaction
times.

The paper is organized as follows. Section \ref{Sec-TCL} contains
a brief account of the TCL projection operator technique and its
application to the model given by the Hamiltonian
(\ref{Hamiltonian}), as well as the derivation of the master
equation governing the dynamics of the reduced density matrix of
the central spin. We compare in Sec.~\ref{Sec-Numerics} the
solutions of this master equation with numerical simulations of
the von Neumann equation corresponding to the Hamiltonian
(\ref{Hamiltonian}). In Sec.~\ref{Sec-ModfiedPicture} we discuss
the performance of an alternative TCL approach that is based on a
modified interaction picture and leads to a simplified master
equation. Finally, we draw our conclusions in
Sec.~\ref{Sec-Conclu}.

\section{TCL master equation} \label{Sec-TCL}

\subsection{Interaction picture}
It is convenient to write the Hamiltonian (\ref{Hamiltonian}) as
$H=H_0+H_I$, where
\begin{equation} \label{H0}
 H_0 = \frac{\omega_0}{2}\sigma_3 + 2\sigma_3K_3
\end{equation}
represents the unperturbed part, and
\begin{equation}\label{H}
 H_I = 2\left(\sigma_+K_- + \sigma_-K_+\right)
\end{equation}
is the interaction Hamiltonian \cite{Coish}. In Eq.~(\ref{H})
$\sigma_{\pm}$ are the raising and lowering operators of the
central spin whereas
\begin{equation}
 K_3 = \frac{1}{2}\sum_{k=1}^N\alpha_k\sigma_3^{k}, \qquad
 K_{\pm} = \sum_{k=1}^N\alpha_k\sigma_{\pm}^{k}.
\end{equation}
In the interaction picture defined by $H_0$ the interaction
Hamiltonian becomes
\begin{equation} \label{HI}
 H_I(t) = \sigma_+B_-(t)+\sigma_-B_+(t),
\end{equation}
where
\begin{equation} \label{B-PM}
 B_{\pm}(t) = 2e^{\mp i\omega_0t}e^{\mp 2iK_3t}K_{\pm}e^{\mp 2iK_3t}.
\end{equation}
Thus the dynamics of the total system's density matrix $\rho(t)$
is governed by the von Neumann equation
\begin{equation}\label{vonneumann}
 \frac{d}{dt}\rho(t) = -i[H_I(t),\rho(t)] \equiv
 \mathcal{L}(t)\rho(t),
\end{equation}
where ${\mathcal{L}}(t)$ denotes the Liouville superoperator
corresponding to the interaction Hamiltonian $H_I(t)$.

\subsection{TCL projection operator approach}
\label{TCL-TECHNIQUE}
The starting point of the projection operator technique is the
introduction of a suitable projection superoperator
${\mathcal{P}}$. This is a positive and trace preserving linear
map that acts on the operators of the total system with the
property of a projection operator, i.~e.
$\mathcal{P}^2=\mathcal{P}$. The superoperator ${\mathcal{P}}$ is
used to project any state $\rho$ of the total system onto its
relevant part ${\mathcal{P}}\rho$, expressing formally the
elimination of the irrelevant degrees of freedom from the full
dynamical description of the underlying model \cite{Breuer1}.

Projection operator techniques are used to derive a closed
equation of motion for the relevant part ${\mathcal{P}}\rho$. A
special variant of these techniques is the time-convolutionless
(TCL) projection operator method. Given an initial state $\rho(0)$
satisfying ${\mathcal{P}}\rho(0)=\rho(0)$, this technique leads to
a time-local first-order master equation for the relevant part of
the form
\begin{equation} \label{TCL0}
 \frac{d}{dt}\mathcal{P}\rho(t) =
 \mathcal{K}(t)\mathcal{P}\rho(t).
\end{equation}
Here, $\mathcal{K}(t)$ is a certain superoperator, representing
the explicitly time-dependent generator of the quantum master
equation for ${\mathcal{P}}\rho$. We note that, like the
corresponding NZ equation, the TCL master equation (\ref{TCL0})
describes all non-Markovian effects although it is local in time.

In practical applications the TCL generator ${\mathcal{K}}(t)$ is
mostly obtained from a perturbation expansion with respect to the
strength of the interaction Hamiltonian,
\begin{equation}\label{TCL-EXPANSION}
 \mathcal{K}(t) = \mathcal{K}_1(t) + \mathcal{K}_2(t) + \ldots
\end{equation}
The first-order contribution is given by
\begin{equation} \label{K-1}
 \mathcal{K}_1(t) = \mathcal{P}\mathcal{L}(t)\mathcal{P},
\end{equation}
while the second-order term takes the form
\begin{equation} \label{K-2}
 \mathcal{K}_2(t) = \int_0^tdt_1 \left[
 \mathcal{P}\mathcal{L}(t)\mathcal{L}(t_1)\mathcal{P} -
 \mathcal{P}\mathcal{L}(t)\mathcal{P}\mathcal{L}(t_1)\mathcal{P}
 \right].
\end{equation}
We remark that this expansion corresponds to an expansion in terms
of the ordered cumulants of the Liouville operator \cite{Breuer1}
${\mathcal{L}}(t)$.

In the present paper we restrict ourselves to the second order and
employ the following projection superoperator,
\begin{equation} \label{CPS1}
 {\mathcal{P}}\rho = \sum_m {\mathrm{Tr}}_B \{ \Pi_m \rho\}
 \otimes \frac{1}{N_m} \Pi_m.
\end{equation}
Here, ${\mathrm{Tr}}_B$ denotes the partial trace over the spin
bath and the $\Pi_m$ are ordinary projection operators acting in
the conventional sense on the Hilbert space of the spin bath. They
project onto the eigenspaces of the 3-component of the bath
angular momentum,
\begin{equation} \label{J3}
 J_3 = \frac{1}{2}\sum_{k=1}^N\sigma_3^k,
\end{equation}
corresponding to the eigenvalues
$m=-\frac{N}{2},\ldots,\frac{N}{2}$. The quantity
\begin{equation}
 N_m = {\mathrm{\Tr}}_B \Pi_m = \binom{N}{\frac{N}{2}+m}
\end{equation}
represents the degree of degeneracy of the eigenvalue $m$ of
$J_3$. Explicitly, we have
\begin{equation}
 \Pi_m = \sum_{\sum m^k = m} |m^1,m^2,\ldots,m^N\rangle
 \langle m^1,m^2,\ldots,m^N|,
\end{equation}
where $m^k=\pm\frac{1}{2}$ denotes the eigenvalue of the $k$-th
bath spin operator $\frac{1}{2}\sigma_3^k$. Obviously, the
projection operators $\Pi_m$ fulfill the relations,
\begin{equation} \label{RELATIONS}
 \Pi_m\Pi_{m'} = \delta_{mm'}\Pi_{m'}, \qquad
 \sum_m\Pi_m=I.
\end{equation}
With the help of Eq.~(\ref{RELATIONS}) it is easy to verify that
the projection superoperator (\ref{CPS1}) is indeed a completely
positive and trace-preserving map that satisfies \cite{CPS}
${\mathcal{P}}^2={\mathcal{P}}$. It projects a given state $\rho$
onto a separable quantum state ${\mathcal{P}}\rho$ which describes
classical correlations between the (unnormalized) system states
\begin{equation}
 \rho_m(t) \equiv \Tr_B\{\Pi_m\rho(t)\}
\end{equation}
and the bath states $\Pi_m/N_m$. The latter represent states of
maximal entropy under the constraint of a given value $m$ for the
total angular momentum. Finally, the reduced density matrix of the
central spin is given by
\begin{equation} \label{RHO-S}
 \rho_S(t) = \Tr_B \, \rho(t) = \sum_m \rho_m(t).
\end{equation}
Thus, the dynamics of the central spin is determined by the
dynamical variables $\rho_m(t)$,
$m=-\frac{N}{2},\ldots,\frac{N}{2}$.

As mentioned already the TCL master equation (\ref{TCL0})
presupposes that the total system's initial state $\rho(0)$
fulfills the condition
\begin{equation} \label{PROD}
 {\mathcal{P}}\rho(0)=\rho(0).
\end{equation}
If this condition is not satisfied one has to add a certain
inhomogeneity to the right-hand side of the TCL master equation
which involves the initial conditions through the complementary
projection ${\mathcal{Q}}\rho(0)=(I-{\mathcal{P}})\rho(0)$. In the
standard applications of the projection operator techniques one
employs a projection superoperator that projects onto an
uncorrelated tensor product state. Condition (\ref{PROD}) then
holds, of course, if and only if the initial state is an
uncorrelated state of the form $\rho(0)=\rho_S\otimes\rho_B$,
where $\rho_S$ is a state of the system and $\rho_B$ is a state of
the bath. However, the projection given by Eq.~(\ref{CPS1})
belongs to the class of {\textit{correlated}} projection
superoperators \cite{CPS,BGM} which projects a given state $\rho$
onto a correlated system-bath state. For this correlated
projection the condition (\ref{PROD}) is satisfied if and only if
the initial state is of the form
\begin{equation} \label{INIT}
 \rho(0) = \sum_m \rho_m(0) \otimes \frac{1}{N_m} \Pi_m.
\end{equation}
Hence, the initial state may contain certain statistical
correlations. A great advantage of the correlated projection
operator technique is therefore given by the fact that it allows
the treatment of correlated initial states by means of a
homogeneous TCL master equation.

\subsection{Deriving the master equation}
For the interaction Hamiltonian (\ref{HI}) the projection operator
(\ref{CPS1}) satisfies $\mathcal{P}\mathcal{L}(t)\mathcal{P}=0$.
Hence, to second order the TCL master equation (\ref{TCL0})
reduces to
\begin{equation}\label{P}
 \frac{d}{dt}{\mathcal{P}}\rho(t) = \int_0^tdt_1
 \mathcal{P}\mathcal{L}(t)\mathcal{L}(t_1)\mathcal{P}\rho(t).
\end{equation}
Taking into account the definition of $\mathcal{P}$ as given by
Eq.~(\ref{CPS1}) and exploiting the properties (\ref{RELATIONS}),
it is possible to convince oneself that Eq.~(\ref{P}) is
equivalent to the following system of coupled differential
equations in the dynamical variables $\rho_m(t)$,
\[
\begin{split}
 \frac{d}{dt}\rho&_m(t) = -\sum_{m'}\int_0^tdt_1 \\
 &\times\Tr_B\left\{
 \Pi_m\left[H_I(t),\left[H_I(t_1),\rho_{m'}(t)\otimes\frac{1}{N_{m'}}\Pi_{m'}
 \right]\right]\right\}.
\end{split}
\]
Evaluating the double commutator we finally get the master
equation
\begin{eqnarray} \label{MASTER-CPS1}
 \frac{d}{d t} \rho_m(t) &=& \int_0^t d\tau \Big\{
 [g_{m+1}(\tau)+g^*_{m+1}(\tau)] \sigma_+\rho_{m+1}(t)\sigma_-
 \nonumber \\
 &~& + [f_{m-1}(\tau)+f^*_{m-1}(\tau)] \sigma_-\rho_{m-1}(t)\sigma_+
 \nonumber \\
 &~& - f_m(\tau) \sigma_+ \sigma_- \rho_{m}(t)
     - f_m^*(\tau) \rho_m(t) \sigma_+ \sigma_- \nonumber \\
 &~& - g_m(\tau) \sigma_- \sigma_+ \rho_{m}(t)
     - g_m^*(\tau) \rho_m(t) \sigma_- \sigma_+ \Big\}. \nonumber \\
\end{eqnarray}
The correlation functions $f_m(\tau)$ and $g_m(\tau)$ are defined
by
\begin{eqnarray}
 f_m(\tau) &=& \left\langle B_-(t) B_+(t_1) \right\rangle_m, \label{F-M} \\
 g_m(\tau) &=& \left\langle B_+(t) B_-(t_1) \right\rangle_m, \label{G-M}
\end{eqnarray}
with $\tau=t-t_1$ and
\begin{equation}
 \langle\mathcal{O}\rangle_m =
 \frac{1}{N_m}\Tr_B\{\mathcal{O}\Pi_m\}.
\end{equation}
Exploiting Eq.~(\ref{B-PM}) we find
\begin{eqnarray}
 f_m(\tau) &=& 4\sum_k \alpha_k^2 \left\langle
 \sigma_-^{k}\sigma_+^{k} e^{i(\omega_0+4K_3+2\alpha_k)\tau}
 \right\rangle_m, \label{f-m} \\
 g_m(\tau) &=& 4\sum_k \alpha_k^2 \left\langle
 \sigma_+^{k}\sigma_-^{k} e^{i(-\omega_0-4K_3+2\alpha_k)\tau}
 \right\rangle_m. \label{g-m}
\end{eqnarray}

\section{Comparison with numerical simulations}
\label{Sec-Numerics}

In this section we compare the dynamics of the density matrix
$\rho_S(t)$ of the central spin obtained by solving the master
equation (\ref{MASTER-CPS1}) with the one deduced directly from
the von Neumann equation (\ref{vonneumann}). In particular we will
consider the dynamics of both the coherences and the populations
of the central spin starting from a fixed initial condition. To
this end, we have carried out numerical simulations of the full
von Neumann equation with mixed initial states for systems with up
to $N=10$ bath spins. Of course, the number of bath spins for
which a direct numerical simulation of the von Neumann equation is
possible is limited by the exponential increase of the dimension
of the underlying Hilbert space. We note that the dimension of the
total state space (the space of density matrices) is given by
$D=2^{2N+2}-1$, which yields $D \approx 4 \cdot 10^6$ for $N=10$.

In the following we assume that the hyperfine coupling constants
are given by
\begin{equation}
 \alpha_k = \alpha_0
 \exp\left[-\left(\frac{k}{k_0}\right)^{n/d}\right],
\end{equation}
where $k_0=N/2$. We choose $n/d=2$ corresponding to a Gaussian
electronic wave function ($n=2$) in one dimension ($d=1$) which is
of relevance in the context of a quantum dot \cite{Coish,Bortz}.
We denote by $A_1$ the mean of the coupling constants $\alpha_k$
and by $A_2$ the respective root mean square,
\begin{equation} \label{DEF-A12}
 A_1 = \frac{1}{N} \sum_k \alpha_k, \qquad
 A_2 = \sqrt{\frac{1}{N} \sum_k \alpha_k^2}.
\end{equation}
The initial state of the total system is taken to be
$\rho(0)=\rho_S(0)\otimes\rho_B(0)$, where $\rho_B(0)=2^{-N}I$
represents an unpolarized infinite temperature state ($I$ denotes
the unit matrix of the spin bath). With this initial state we have
$\rho_m(0)=2^{-N}N_m\rho_S(0)$. We emphasize that the present
technique also allows the treatment of polarized and of correlated
initial states (see Sec.~\ref{TCL-TECHNIQUE}).

\subsection{Coherences}
The coherence of the central spin is defined by
$\tilde{C}(t)=\langle +|\rho_S(t)|-\rangle$, where $|\pm\rangle$
denote the eigenstates of $\sigma_3$. According to
Eq.~(\ref{RHO-S}) we have
\begin{equation}
 \tilde{C}(t) = \sum_m \tilde{C}_m(t)
 = \sum_m \langle +|\rho_m(t)|-\rangle.
\end{equation}
Starting from the master equation (\ref{MASTER-CPS1}) we have
\begin{equation}\label{TCLcoh}
 \frac{d}{dt}\tilde{C}_m(t) = -\int_0^t d\tau
 [f_m(\tau)+g^*_m(\tau)] \tilde{C}_m(t)
\end{equation}
with the obvious solution
\begin{equation}
 \tilde{C}_m(t) = \tilde{C}_m(0) e^{-\Lambda^{\mathrm{coh}}_m(t)},
\end{equation}
where
\begin{equation} \label{INT-1}
 \Lambda^{\mathrm{coh}}_m(t) = \int_0^t dt_1 \int_0^{t_1} dt_2
 \left[f_m(t_2)+g_m^*(t_2)\right].
\end{equation}
Let's now remember that the master equation (\ref{MASTER-CPS1})
has been written in the interaction picture with respect to the
Hamiltonian (\ref{H0}). As usual, we will represent our results in
the interaction picture with respect to the free Hamiltonian
$\frac{\omega_0}{2}\sigma_3$, that is in the rotating frame of the
central spin. In order to do this we have to use the replacement
\[
 \sum_m \rho_m(t) \otimes \frac{\Pi_m}{N_m} \rightarrow
 \sum_m e^{-2i\sigma_3K_3t} \rho_m(t) \otimes \frac{\Pi_m}{N_m} e^{2i\sigma_3K_3t}.
\]
It is immediate to observe that under this transformation the
populations remain unchanged, while the coherences must be
multiplied by the factor $\left\langle e^{-4iK_3t}
\right\rangle_m$. Hence, the coherence $C(t)$ in the rotating
frame of the central spin is found to be
\begin{equation} \label{coh}
 C(t) = C(0) \sum_m \frac{N_m}{2^N} \left\langle e^{-4iK_3t} \right\rangle_m
 e^{-\Lambda^{\mathrm{coh}}_m(t)}.
\end{equation}

\begin{figure}[htb]
\begin{center}
\includegraphics[width=0.8\linewidth]{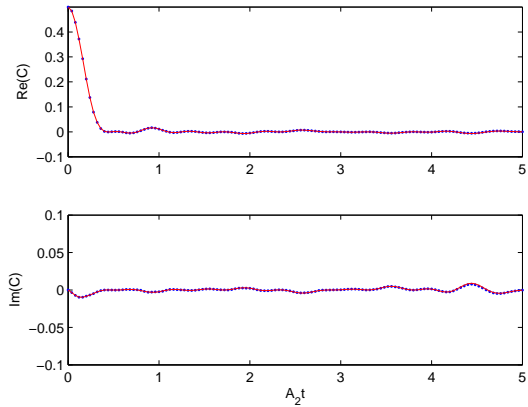}
\caption{(Color online) Real and imaginary part of the coherence
of the central spin for $\alpha_0/\omega_0=0.01$ and $N=10$ bath
spins. Blue dots: Numerical simulation. Red line: TCL
approximation according to Eq.~(\ref{coh}).} \label{fig1}
\end{center}
\end{figure}

In Fig.~\ref{fig1} we compare the result of the TCL approximation
given by Eq.~(\ref{coh}) with numerical simulations of the von
Neumann equation (\ref{vonneumann}) supposing that the central
system is initially prepared in the superposition state
$\frac{1}{\sqrt{2}}(|+\rangle+|-\rangle)$. The correlation
functions (\ref{f-m}) and (\ref{g-m}), and the mean value
$\left\langle e^{-4iK_3t} \right\rangle_m$ have been calculated
numerically. As it is evident from the Figures the agreement
between the TCL result and the numerical solution is excellent in
the perturbation regime, even for long integration times.

\subsection{Populations}
The populations $P_{\pm}(t)=\langle\pm|\rho_S(t)|\pm\rangle$ of
the central spin are given by
\begin{equation}
 P_{\pm}(t) = \sum_m P_m^{\pm}(t) = \sum_m \langle\pm|\rho_m(t)|\pm\rangle.
\end{equation}
The master equation (\ref{MASTER-CPS1}) leads to a system of
coupled equations,
\begin{eqnarray}
 \frac{d}{d t} P_m^+(t) &=& \int_0^t d\tau
 [g_{m+1}(\tau)+g^*_{m+1}(\tau)] P_{m+1}^-(t) \nonumber \\
 &~& - [f_{m}(\tau)+f^*_{m}(\tau)] P_{m}^+(t), \\
 \frac{d}{d t} P_m^-(t) &=& \int_0^t d\tau
 [f_{m-1}(\tau)+f^*_{m-1}(\tau)] P_{m-1}^+(t) \nonumber\\
 &~& - [g_{m}(\tau)+g^*_{m}(\tau)] P_{m}^-(t).
\end{eqnarray}
To solve these equations we employ the relation
\begin{equation} \label{CONS-P}
 \frac{d}{d t}\left[ P_m^+(t) + P_{m+1}^-(t) \right] = 0,
\end{equation}
which expresses the conservation of the 3-component of the total
spin angular momentum. Using the initial condition $P_+(0)=1$ we
obtain
\begin{equation} \label{P+}
 P_+(t) = \sum_m \frac{N_m}{2^N} e^{-\Lambda^{\mathrm{pop}}_m(t)} \left[
 1 + \int_0^t dt_1 e^{\Lambda^{\mathrm{pop}}_m(t_1)} \mu_m(t_1) \right],
\end{equation}
where
\begin{equation} \label{INT-2}
 \Lambda^{\mathrm{pop}}_m(t) = 2{\mathrm{Re}}\int_0^t dt_1 \int_0^{t_1} dt_2
 \left[g_{m+1}(t_2)+f_m(t_2)\right],
\end{equation}
and
\begin{equation} \label{INT-3}
 \mu_m(t) = 2{\mathrm{Re}}\int_0^t d\tau \, g_{m+1}(\tau).
\end{equation}

\begin{figure}[htb]
\begin{center}
\includegraphics[width=0.8\linewidth]{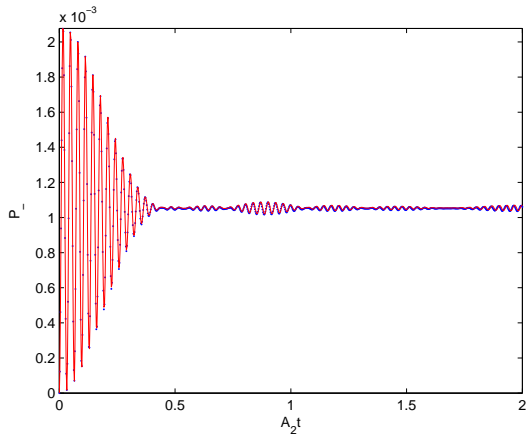}
\caption{(Color online) Population of the central spin for
$\alpha_0/\omega_0=0.01$ and $N=10$ bath spins. Blue dots:
Numerical simulation. Red line: TCL approximation according to
Eq.~(\ref{P+}).} \label{fig2}
\end{center}
\end{figure}

\begin{figure}[htb]
\begin{center}
\includegraphics[width=0.8\linewidth]{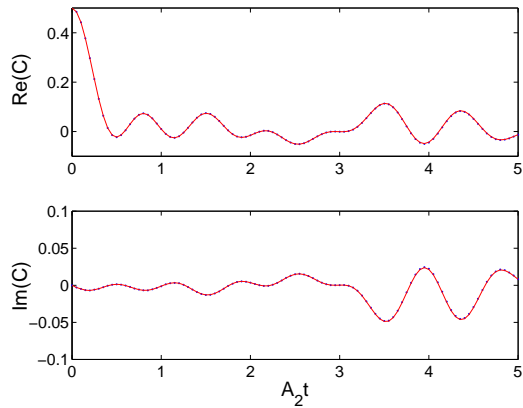}
\caption{(Color online) Real and imaginary part of the coherence
of the central spin for $\alpha_0/\omega_0=0.01$ and $N=6$ bath
spins. Blue dots: Numerical simulation. Red line: TCL
approximation according to Eq.~(\ref{coh}).} \label{fig3}
\end{center}
\end{figure}

\begin{figure}[htb]
\begin{center}
\includegraphics[width=0.8\linewidth]{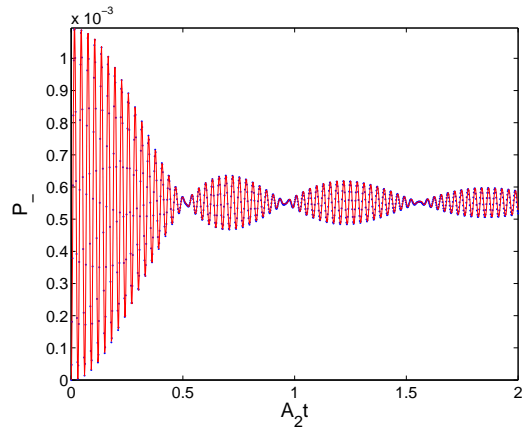}
\caption{(Color online) Population of the central spin for
$\alpha_0/\omega_0=0.01$ and $N=6$ bath spins. Blue dots:
Numerical simulation. Red line: TCL approximation according to
Eq.~(\ref{P+}).} \label{fig4}
\end{center}
\end{figure}

The comparison of the TCL result (\ref{P+}) with the numerical
simulation is shown in Fig.~\ref{fig2}. We again observe a very
good agreement of the TCL approximation with the exact dynamics
for short and also for long interaction times. Due to the
exponential increase of the numerical effort with increasing $N$
we can treat only a relatively small number of bath spins. Thus we
have an open quantum system, the central spin, which is coupled to
a relatively small environment, a finite system of bath spins. For
this reason, the conventional techniques used in the theory of
open systems to derive a Markovian master equation are not
applicable because they are usually based on an effectively
infinite environment with a continuum of bath modes. However, the
TCL approximation scheme developed here does not require that $N$
be large. The master equation (\ref{MASTER-CPS1}) is therefore
valid also for a very small number of bath spins. We illustrate
this point in Figs.~\ref{fig3} and \ref{fig4} which show the
dynamics of the coherences and the populations for $N=6$ bath
spins.

\begin{figure}[htb]
\begin{center}
\includegraphics[width=0.8\linewidth]{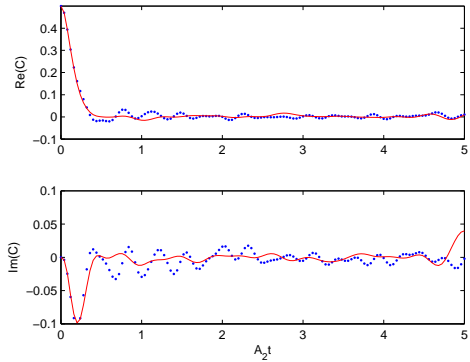}
\caption{(Color online) Real and imaginary part of the coherence
of the central spin for $\alpha_0/\omega_0=0.1$ and $N=10$ bath
spins. Blue dots: Numerical simulation. Red line: TCL
approximation according to Eq.~(\ref{coh}).} \label{fig5}
\end{center}
\end{figure}

\begin{figure}[htb]
\begin{center}
\includegraphics[width=0.8\linewidth]{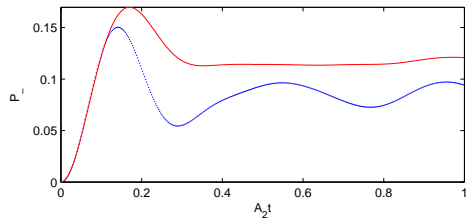}
\caption{(Color online) Population of the central spin for
$\alpha_0/\omega_0=0.1$ and $N=10$ bath spins. Blue dots:
Numerical simulation. Red line: TCL approximation according to
Eq.~(\ref{P+}).} \label{fig6}
\end{center}
\end{figure}

The master equation (\ref{MASTER-CPS1}) has been obtained from the
second order of the TCL perturbation expansion. Of course, for
much stronger system-bath couplings the second-order result fails,
indicating the relevance of cumulants of higher-order. To
illustrate this point we have increased $\alpha_0$ by a factor of
10. The result for the coherences and the populations is depicted
in Figs.~\ref{fig5} and \ref{fig6}, respectively. We observe that
the short-time behavior is still correctly reproduced by the
second-order, while there are large deviations for longer times.
We conclude from our numerical simulations that the second order
of the TCL scheme yields a good agreement with the exact dynamics
for couplings up to the order of $\alpha_0/\omega_0 \sim 10^{-2}$.
Note however that the decay of the coherence $C(t)$ and the
corresponding decoherence time are very well reproduced even for
much larger couplings as can be seen from Figs.~\ref{fig5}.

\section{Modified interaction picture} \label{Sec-ModfiedPicture}
A certain disadvantage of the perturbation scheme used in
Sec.~\ref{Sec-TCL} consists in the fact that the time integrals
over the correlations functions in Eqs.~(\ref{INT-1}) and
(\ref{P+})-(\ref{INT-3}) are generally difficult to calculate and
have to be determined numerically. To avoid the appearance of
these expressions and to obtain a simpler approximation scheme we
employ a modification of the interaction picture Hamiltonian. To
this end, we write the Hamiltonian (\ref{Hamiltonian}) again as
$H=H_0+H_I$, where now the unperturbed part is given by
\begin{equation} \label{H0-MOD}
 H_0 = \frac{\omega_0}{2}\sigma_3 + 2A_2\sigma_3J_3,
\end{equation}
and
\begin{equation}
 H_I = 2\sigma_3 \left( K_3-A_2J_3 \right)
 + 2\left(\sigma_+K_- + \sigma_-K_+\right)
\end{equation}
represents the interaction Hamiltonian. By contrast to the
interaction picture of Sec.~\ref{Sec-TCL}, here the diagonal term
$2\sigma_3K_3$ of the Hamiltonian (\ref{Hamiltonian}) is not
completely removed from the interaction, but we subtract only the
term $2A_2\sigma_3J_3$ involving an effective coupling constant
$A_2$ that is equal to the root mean square of the $\alpha_k$ [see
Eqs.~(\ref{J3}) and (\ref{DEF-A12})]. Hence, the interaction
picture Hamiltonian now takes the form
\begin{equation} \label{HI-MOD}
 H_I(t) = 2\sigma_3 \left( K_3-A_2J_3 \right) + \sigma_+B_-(t)+\sigma_-B_+(t),
\end{equation}
where
\begin{equation}
 B_{\pm}(t) = 2e^{\mp i\omega_0t}e^{\mp 2iA_2J_3t}K_{\pm}e^{\mp 2iA_2J_3t}.
\end{equation}
The corresponding Liouville operator will again be denoted by
${\mathcal{L}}(t)$.

An important point of the new interaction picture is that for the
Hamiltonian (\ref{HI-MOD}) and the projection (\ref{CPS1}) the
first order term $\mathcal{P}\mathcal{L}(t)\mathcal{P}$ does not
vanish. Hence, one has to use the full expressions (\ref{K-1}) and
(\ref{K-2}) for the second-order TCL generator. However, the great
advantage of the present procedure is the fact that the
correlation functions (\ref{F-M}) and (\ref{G-M}) take on a very
simple form,
\begin{eqnarray}
 f_m(\tau) &=& B_+(m) e^{i\Omega_+(m)\tau}, \\
 g_m(\tau) &=& B_-(m) e^{i\Omega_-(m)\tau}, \
\end{eqnarray}
where
\[
 \Omega_{\pm}(m)=\pm\omega_0+4A_2\left(\pm m + \frac{1}{2}\right)
\]
and
\[
 B_{\pm}(m) = 4A_2^2\left(\frac{N}{2}\mp m\right).
\]
With the help of these expressions we find the master equation
\begin{eqnarray} \label{MASTER-CPS2}
 \frac{d}{d t} \rho_m(t) &=& 2im(A_2-A_1) [\sigma_3,\rho_m(t)] \nonumber \\
 &~& - \frac{N^2-4m^2}{N-1}(A_2^2-A_1^2) t
 [\sigma_3,[\sigma_3,\rho_m(t)]] \nonumber \\
 &~& + \int_0^t d\tau \nonumber \\
 &\times& \Big\{
 B_-(m+1) 2\cos[\Omega_+(m)\tau] \sigma_+\rho_{m+1}(t)\sigma_-
 \nonumber \\
 &~& + B_+(m-1) 2\cos[\Omega_-(m)\tau] \sigma_-\rho_{m-1}(t)\sigma_+
 \nonumber \\
 &~& - B_+(m) e^{i\Omega_+(m)\tau} \sigma_+ \sigma_- \rho_{m}(t)
 \nonumber \\
 &~& - B_+(m) e^{-i\Omega_+(m)\tau} \rho_{m}(t) \sigma_+ \sigma_-
 \nonumber \\
 &~& - B_-(m) e^{i\Omega_-(m)\tau} \sigma_- \sigma_+ \rho_{m}(t)
 \nonumber \\
 &~& - B_-(m) e^{-i\Omega_-(m)\tau} \rho_m(t) \sigma_- \sigma_+
 \Big\}.
\end{eqnarray}
In the special case of uniform couplings
($\alpha_k={\mbox{const}}$) we have $A_1=A_2$. Equation
(\ref{MASTER-CPS2}) then reduces to the master equation derived in
Ref. $17$.

\begin{figure}[htb]
\begin{center}
\includegraphics[width=0.8\linewidth]{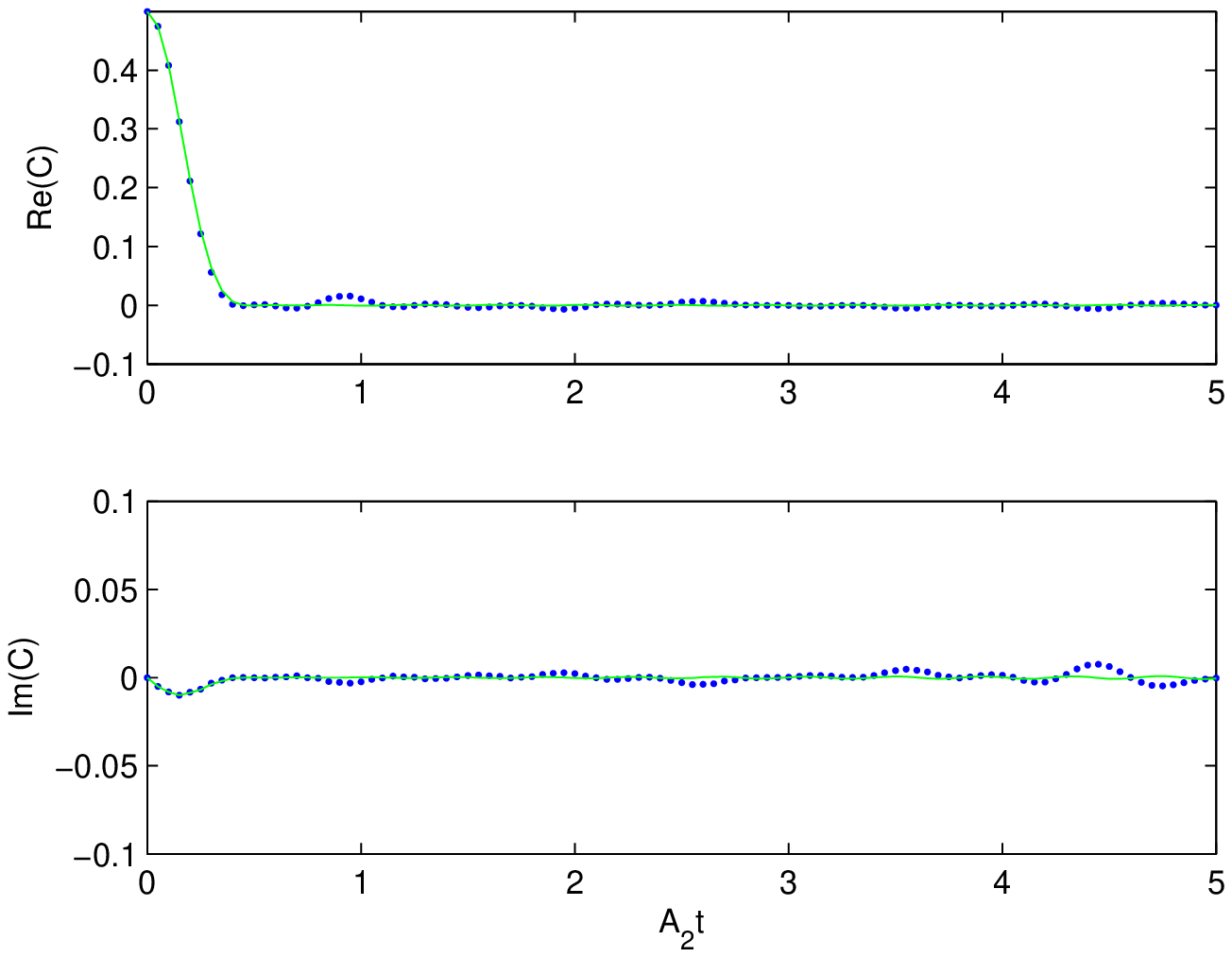}
\caption{(Color online) Real and imaginary part of the coherence
of the central spin for $\alpha_0/\omega_0=0.01$ and $N=10$. Blue
dots: numerical simulation. Green line: TCL approximation
according to Eq.~(\ref{coh-mod}).} \label{fig7}
\end{center}
\end{figure}

\begin{figure}[htb]
\begin{center}
\includegraphics[width=0.8\linewidth]{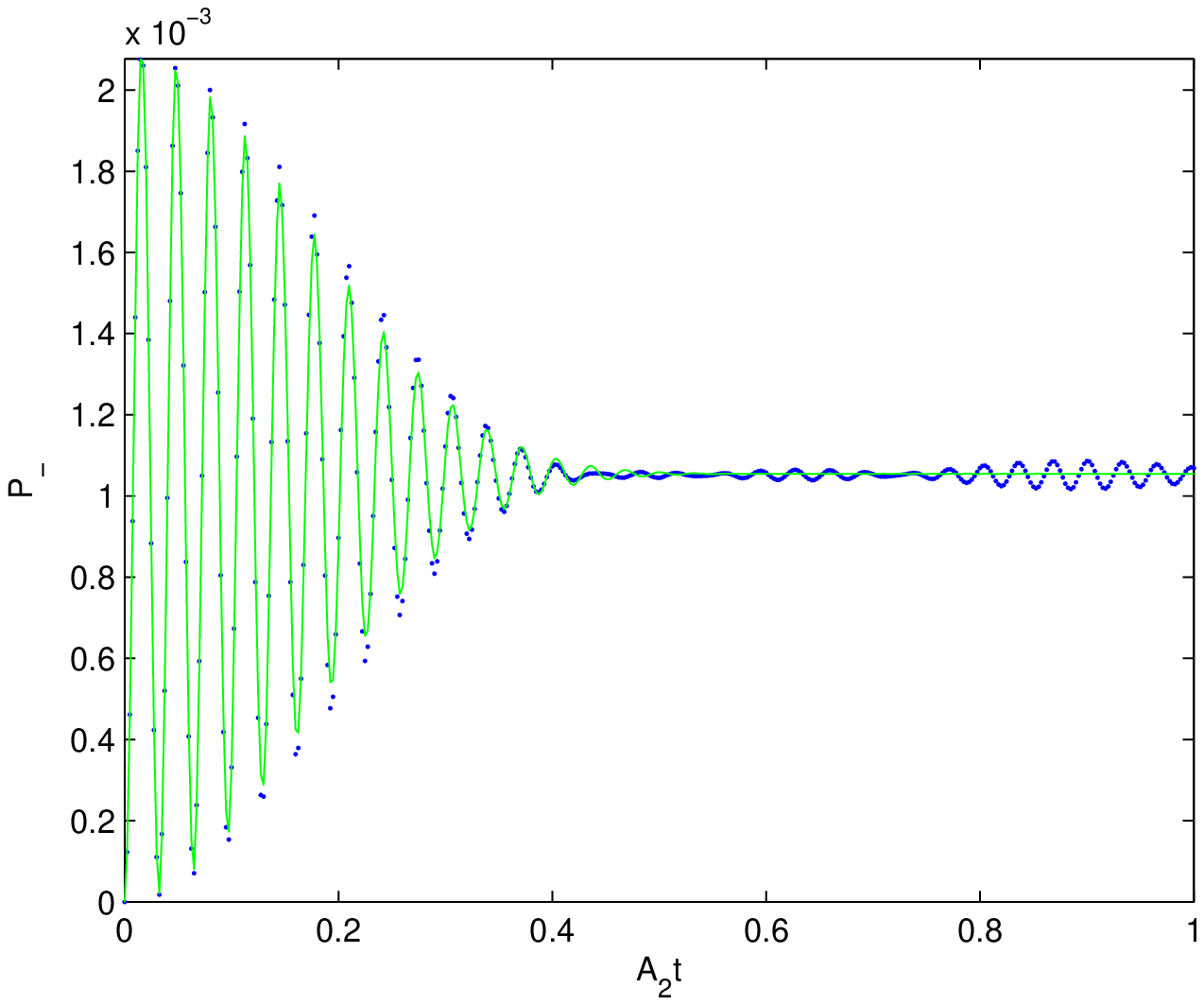}
\caption{(Color online) Population of the central spin for
$\alpha_0/\omega_0=0.01$ and $N=10$. Blue dots: Numerical
simulation. Green line: TCL approximation according to
Eq.~(\ref{P+-mod}).} \label{fig8}
\end{center}
\end{figure}

\begin{figure}[htb]
\begin{center}
\includegraphics[width=0.8\linewidth]{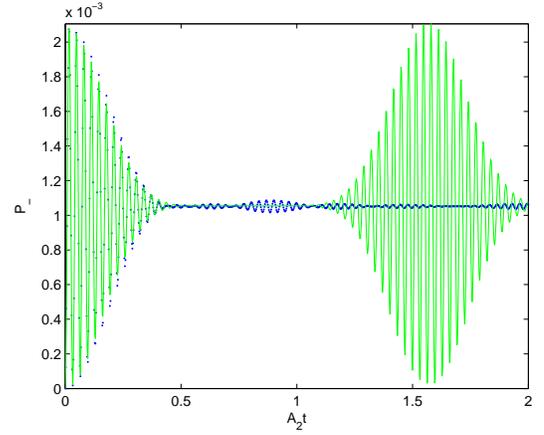}
\caption{(Color online) The same as Fig.~\ref{fig8} for a longer
interaction time.} \label{fig9}
\end{center}
\end{figure}

The master equation (\ref{MASTER-CPS2}) can again be solved
analytically. The procedure is similar to the one outlined in
Sec.~\ref{Sec-TCL}. We find the coherences,
\begin{equation} \label{coh-mod}
 C(t) = C(0) \sum_m \frac{N_m}{2^N} e^{-\Lambda^{\mathrm{coh}}_m(t)},
\end{equation}
where
\begin{eqnarray}
 \Lambda^{\mathrm{coh}}_m(t) &=& 4iA_1mt
 +2\frac{N^2-4m^2}{N-1}(A_2^2-A_1^2)t^2 \nonumber \\
 && + \frac{B_+(m)}{\Omega_+^2(m)} \left[1-e^{i \Omega_+(m)t}\right]
 \nonumber \\
 && + \frac{B_-(m)}{\Omega_-^2(m)} \left[1-e^{-i \Omega_-(m)t}\right]
 \nonumber \\
 && + it \left[ \frac{B_+(m)}{\Omega_+(m)} - \frac{B_-(m)}{\Omega_-(m)}
 \right],
\end{eqnarray}
and the populations,
\begin{equation} \label{P+-mod}
 P_+(t) = \sum_m \frac{N_m}{2^N} \left[
 \frac{\frac{N}{2}+m+1}{N+1} + \frac{\frac{N}{2}-m}{N+1}
 e^{-\Lambda^{\mathrm{pop}}_m(t)}
 \right],
\end{equation}
where
\begin{equation}
 \Lambda^{\mathrm{pop}}_m(t) = \frac{8A_2^2(N+1)}{\Omega_+^2(m)}
 \left( 1-\cos[\Omega_+(m)t] \right).
\end{equation}

Involving only a sum over the quantum number $m$, the expressions
(\ref{coh-mod}) and (\ref{P+-mod}) can be evaluated numerically in
a very efficient way. The comparison with the numerical simulation
of the von Neumann equation demonstrates that also the TCL
approach with the modified interaction picture yields a good
agreement. The decoherence (see Fig.~\ref{fig7}) as well as the
oscillations and the decay of the populations (see
Fig.~\ref{fig8}) are very well reproduced by the simplified
scheme. For longer interaction times the result (\ref{P+-mod})
leads to revivals of the populations (see Fig.~\ref{fig9}) which
are due to the commensurability of the frequencies $\Omega_+(m)$.
We stress that these revivals are neither present in the exact
solution nor in the TCL approximation (\ref{P+}). Apart from these
revivals the simplified approximation given by the master equation
(\ref{MASTER-CPS2}) thus provides an accurate description of the
decoherence and of the oscillating decay of the populations.

Finally we investigate the limit of a large number $N$ of bath
spins. For small values of the quantity $\beta \equiv
2\sqrt{N}A_2/\omega_0$ and large $N$ the result (\ref{P+-mod}) can
be approximated by
\begin{equation} \label{P-approx}
 P_+(t) \approx 1 - \beta^2 \left[
 1 - e^{-2NA_2^2 t^2} \cos\omega_0 t \right].
\end{equation}
To obtain this expression one first expands the exponential in
Eq.~(\ref{P+-mod}) for small $\Lambda_m^{\mathrm{pop}}$ and
carries out the summation over $m$. Equation (\ref{P-approx})
provides a good approximation for a fixed $\beta \ll 1$ even for
moderate $N$-values. For example, the case of $N=10$ bath spins
with $\alpha_0/\omega_0=0.01$ investigated above corresponds to
$\beta = 0.03$. For this value of $\beta$, we find that
Eq.~(\ref{P-approx}) yields a good agreement for all $N$ larger
than about 10.

\section{Conclusions} \label{Sec-Conclu}
The appearance of a retarded memory kernel in the equations of
motion, e.~g. in the Nakajima-Zwanzig equation, is often regarded
as \textit{the} characteristic feature of non-Markovian quantum
processes. However, applications of the time-convolutionless
projection operator technique show that strong non-Markovian
behavior of open quantum systems can often be described by
time-local master equations with an explicitly time-dependent
generator. Although there is no theory of non-Markovian dynamics
which allows a general assessment of the NZ and the TCL scheme,
experience shows that in many physical systems the degree of
accuracy achieved by both methods is of the same order of
magnitude. In such cases the TCL method is clearly to be preferred
because it only requires solving a time-local master equation.
This does not mean that the TCL technique is always better than
the NZ technique. The performance of these perturbation schemes
strongly depends on the details of the system under investigation
and on the chosen projection superoperator. There are examples of
physical systems for which either the NZ or the TCL approach
yields the exact result already in lowest order of perturbation
theory\cite{Fischer,DIFBAL}.

In the present paper we have demonstrated the feasibility of the
TCL approach in the case of a nontrivial model describing a
central spin coupled to a spin bath through nonuniform Heisenberg
interaction. We have shown that the method indeed works in the
strong field limit and provides a good approximation of the short
and the long-time behavior of the coherences and the populations
of the central spin. In addition, we have developed a TCL master
equation that is based on a modified interaction picture and leads
to a compact analytical solution for the central spin's density
matrix which allows an efficient numerical computation.

A possible approach to moderate or strong couplings is the
analysis of the ordered cumulants of higher orders in the TCL
expansion. However, it seems that a more efficient strategy
consists in the construction of more suitable correlated
projection superoperators. An example of this strategy is
discussed in Ref.~\onlinecite{Fischer}, where a certain correlated
projection has been constructed for which the second order
Nakajima-Zwanzig master equation leads to the exact dynamics of
the population for the Hamiltonian (\ref{Hamiltonian}) with
uniform couplings. It is of great relevance to develop possible
extensions of this strategy to the nonuniform spin bath model
analyzed here.

\begin{acknowledgments}
AM (AN) acknowledges partial support by MIUR project II04C0E3F3
(II04C1AF4E) \textit{Collaborazioni Interuniversitarie ed
Internazionali tipologia C}. H.P.B. gratefully acknowledges
financial support from Hanse- Wissenschaftskolleg, Delmenhorst.
\end{acknowledgments}


\begin{thebibliography}{19}

\bibitem{KUBO} R. Kubo, M. Toda, and N. Hashitsume, \textit{Statistical Physics II.
Nonequilibrium Statistical Mechanics}, (Springer, Berlin, 1991).

\bibitem{Breuer1} H.-P. Breuer and F. Petruccione, \textit{The Theory of Open
Quantum Systems}, (Oxford University Press, Oxford, 2007).

\bibitem{NAKAJIMA} S. Nakajima, Progr. Theor. Phys. \textbf{20}, 948 (1958).

\bibitem{ZWANZIG} R. Zwanzig, J. Chem. Phys. \textbf{33}, 1338 (1960).

\bibitem{SHIBATA} S. Chaturvedi and F. Shibata, Z. Phys. B \textbf{35}, 297 (1979).

\bibitem{DECOHERENCE} H. P. Breuer, B. Kappler and F. Petruccione, Ann.
Phys. (N.Y.) \textbf{291}, 36 (2001).

\bibitem{AHN02} D. Ahn, J. Lee, M. S. Kim, and S. W. Hwang,
Phys. Rev. A \textbf{66}, 012302 (2002).

\bibitem{BGM} H. P. Breuer, J. Gemmer, and M. Michel,
Phys. Rev. E \textbf{73}, 016139 (2006).

\bibitem{Schliemann} J. Schliemann, A. Khaetskii and D. Loss,
J. Phys.: Condens. Matter \textbf{15}, R1809 (2003).

\bibitem{Coish} W. A. Coish and D. Loss, Phys. Rev. B \textbf{70}, 195340 (2004).

\bibitem{Khaetskii} A. V. Khaetskii, D. Loss and L. Glazman, Phys Rev. Lett.
\textbf{88}, 186802 (2002).

\bibitem{Gaudin} M. Gaudin, J. Phys. (France) \textbf{37}, 1087
(1976).

\bibitem{Bortz} M. Bortz and J. Stolze, Phys. Rev. B \textbf{76}, 014304 (2007).

\bibitem{Harmon} K. A. Al-Hassanieh, V. V. Dobrovitski,  E. Dagotto, and B. N.
Harmon, Phys. Rev. Lett. \textbf{97}, 037204 (2006).

\bibitem{DeRaedt} V. V. Dobrovitski and H. A. De Raedt, Phys. Rev.
E \textbf{67}, 056702 (2003).

\bibitem{BBP} H. P. Breuer, D. Burgarth and F. Petruccione, Phys. Rev. B \textbf{70},
045323 (2004).

\bibitem{Fischer}
J. Fisher and H.-P. Breuer, Phys. Rev. A \textbf{76}, 052119
(2007).

\bibitem{LIDAR07} H. Krovi, O. Oreshkov, M. Ryazanov, and D. A. Lidar,
Phys. Rev. A \textbf{76}, 052117 (2007).

\bibitem{CPS} H.-P. Breuer, Phys. Rev. A \textbf{75}, 022103 (2007).

\bibitem{DIFBAL} R. Steinigeweg, H.-P. Breuer and J. Gemmer, Phys.
Rev. Lett. \textbf{99}, 150601 (2007).


\end{thebibliography}
\end{document}